\documentclass[referee]{aa} 
\usepackage{txfonts}
\usepackage{graphicx,epsfig}

\newcommand{\Mearth}{$M_\oplus$}

\begin{document}
   \title{Elemental abundances and minimum mass of heavy elements in the envelope of HD 189733b}

   \author{Olivier~Mousis
          \inst{1,}\inst{2,}\inst{5}
          \and
          Jonathan~I.~Lunine\inst{1}
          \and 
          Giovanna~Tinetti\inst{3,}\inst{5}
          \and
          Caitlin~A.~Griffith\inst{1}
          \and
          Adam~P.~Showman\inst{1}
          \and
          Yann~Alibert\inst{2}
          \and
          Jean-Philippe~Beaulieu\inst{4,}\inst{5}
       	}

\institute{Lunar and Planetary Laboratory, University of Arizona, Tucson, AZ, USA \and Institut UTINAM, CNRS-UMR 6213, Observatoire de  Besan\c{c}on, Universit{\'e} de Franche-Comt{\'e}, Besan\c{c}on, France \and Department of Physics and Astronomy, University College London, London, UK \and Institut d'astrophysique de Paris, CNRS-UMR 7095, Universit\'e Pierre \& Marie Curie, France \and The HOLMES collaboration}
\offprints{O. Mousis, \\   e-mail: olivier.mousis@obs-besancon.fr}

   \date{Received; accepted}

 
  \abstract
   {Oxygen (O) and carbon (C) have been inferred recently to be subsolar in abundance from spectra of the atmosphere of the transiting hot Jupiter HD 189733b. Yet, the mass and radius of the planet coupled with structure models indicate a strongly supersolar abundance of heavy elements in the interior of this object.}
   {Here we explore the discrepancy between the large amount of heavy elements suspected in the planet's interior and the paucity of volatiles measured in its atmosphere.}
   {We describe the formation sequence of the icy planetesimals formed beyond the snow line of the protoplanetary disk and calculate the composition of ices ultimately accreted in the envelope of HD 189733b on its migration pathway. This allows us to reproduce the observed volatile abundances by adjusting the mass of ices vaporized in the envelope.}
   {The predicted elemental mixing ratios should be 0.15--0.3 times solar in the envelope of HD 189733b if they are fitted to the recent O and C determinations. However, our fit to the minimum mass of heavy elements predicted by internal structure models gives elemental abundances that are 1.2--2.4 times oversolar in the envelope of HD189733b.}
   {We propose that the most likely cause of this discrepancy is irradiation from the central star leading to development of a radiative zone in the planet's outer envelope which would induce gravitational settling of elements. Hence, all strongly irradiated extrasolar planets should present subsolar abundances of volatiles. We finally predict that the abundances of nitrogen (N), sulfur (S) and phosphorus (P) are of $\sim$ $2.8 \times 10^{-5}$, $5.3 \times 10^{-6}$ and $1.8 \times 10^{-7}$ relative to H$_2$, respectively in the atmosphere of HD 189733b.}

   \keywords{planetary systems -- planetary systems: formation -- planetary systems: protoplanetary disks.}

   \maketitle
%

\section{Introduction}

HD 189733b is a transiting hot Jupiter ($M =1.15 \pm 0.04 M_J$) orbiting a bright (V = 7.7) and close (d = 19 pc) K2V stellar primary at the distance of 0.03 AU (Bouchy et al. 2005). Observations dedicated to the characterization of the atmosphere of HD 189733b during secondary eclipses revealed the presence of H$_2$O (Tinetti et al. 2007), CH$_4$ (Swain et al. 2008) and CO (Charbonneau et al. 2008). The dayside spectrum of HD 189733b has also been analysed at near-IR wavelenghts by Swain et al. (2009a) (hereafter S09a) who confirmed the presence of the previously discovered molecules and announced the detection of CO$_2$. Emission spectra probe both atmospheric temperature and composition. Initial estimates of the composition, found to depend on the assumed temperature profile, were derived from a range of profiles, and are shown in Table \ref{ratios}  (S09a). 

In the solar system, the atmospheric compositions of Jupiter and Saturn provide important constraints on the formation processes of these planets (Alibert et al. 2005a, 2005b), on their internal structures and the thermodynamic conditions that existed in the primitive nebula at the epoch of formation of their solid and gaseous building blocks (Hersant et al. 2008; Mousis et al. 2009). The observed supersolar abundances of volatiles in Jupiter and Saturn can be explained by the accretion into the envelopes of the two forming planets of planetesimals composed of a mix of rock, clathrate hydrates (hereafter clathrates) and pure ices formed from a gas phase of solar composition (Mousis et al. 2009), in a way that is consistent with the so-called core-accretion formation model (Pollack et al. 1996).

\begin{table}
\caption[]{Elemental and molecular abundances measured in HD 189733b (from S09a).}
\begin{center}
\begin{tabular}{lccc}
\hline
\hline
\noalign{\smallskip}
Species X 	&  (X/H$_2$)   				& Species X 		&  (X/H$_2$) 				\\	
\noalign{\smallskip}
\hline
\noalign{\smallskip}
O			& $1.1 - 4 \times 10^{-4}$		& CO			& $1 - 3    \times 10^{-4}$		\\
C			& $1 - 3 \times 10^{-4}$		& CO$_2$		& $0.1 - 1 \times 10^{-6}$		\\
H$_2$O		& $0.1 - 1 \times 10^{-4}$		& CH$_4$		& $ \le 1 \times 10^{-7}$		\\

\hline
\end{tabular}
\end{center}
\label{ratios}
\end{table}

In the case of HD 189733b, the puzzling feature of the O and C determinations by S09a is that they have been inferred to be subsolar in the atmosphere while internal structure models indicate a strongly supersolar abundance of heavy elements in the interior of this object (Guillot 2008). A broader range of abundances are indicated by works that consider temperature profiles not included in the initial discoveries of molecules in extrasolar planetary atmospheres (Madhusudhan \& Seager 2009; Swain et al. 2009b) and current analyses of both transmission and emission data aim to better constrain the elemental abundances. Although consensus has not yet been reached on the metallicity of HD 189733b, the possibility that this -- or potentially other -- hot Jupiters have subsolar metallicity raises the interesting theoretical question of whether and how a giant planet may achieve subsolar metallicity during its formation.  Here, we address this theoretical question. We use our model of the formation sequence of ices in protoplanetary disks to determine the amount of heavy elements needed to account for the recent observations of O and C abundances in the atmosphere of HD 189733b by S09a. We show that the required amount of accreted ices in HD 189733b is much smaller than the minimum mass of heavy elements predicted by internal structure models. We then investigate the reasons that could explain this discrepancy and conclude that the most likely cause would be the development of a radiative zone in the outer part of the planet's envelope which, combined with gravity, would induce gravitational settling of elements. We finally predict the N, S and P abundances in the outer part of HD 189733b's atmosphere.

\section{Delivery of heavy elements to proto-HD 189733b}

Close-in giant planets are thought to have formed in the cold outer region of protoplanetary disks and migrated inwards until they stopped at very small orbital radii to the star (Goldreich \& Tremaine 1980; Lin et al. 1996; Fogg \& Nelson 2005, 2007; Mandell et al. 2007). In this context, we follow the core-accretion model to describe the formation of HD189733b in which a solid core is first formed by collisional accretion of planetesimals in the outer part of the disk (defined to be beyond the ice condensation, or `snow" line). In a second phase, when the mass of this core becomes large enough, rapid accretion of gas and gas-coupled solids is triggered, leading to the formation of a gas giant planet (Pollack et al. 1996; Alibert et al. 2005a, 2005b). Building blocks accreted by proto-HD 189733b may have formed all along its radial migration pathway in the protoplanetary disk. However, we assume here that only the planetesimals produced beyond the snow line, i.e. those possessing a significant fraction of volatiles, materially affected the observed O and C abundances due to their vaporization when they entered the envelope of the planet.

\section{Formation sequence of icy planetesimals}

We describe the formation sequence of the different ices formed beyond the snow line of the protoplanetary disk. For this purpose, we use the solar nebula model employed by Mousis et al. (2009) for calculating the condensation sequence of the different volatiles in the feeding zone of HD 189733b. We refer the reader to the work of Alibert et al. (2005c) for a full description of this model of an accretion disk. Once formed, these ices will add to the composition of the planetesimals accreted by the growing planet on its migration pathway. Our calculation of the composition of these ices is made easier  by the fact that it does not depend on their formation location or the adopted disk thermodynamic conditions (Marboeuf et al. 2008). On the other hand, it does depend significantly on the initial elemental composition of the disk's gas phase (Marboeuf et al. 2008), which is assumed to be solar in the present case, in agreement with the measured parent star metallicity ([Fe/H] = -0.03 $\pm$ 0.04; Bouchy et al. 2005). This implies that the planetesimals accreted by proto-HD 189733b on its migration pathway are similar to each other in composition, provided that the gas phase composition of the disk remains unchanged.

\begin{table}
\caption[]{Elemental and molecular abundances in the protoplanetary disk.}
\begin{center}
\begin{tabular}{lccc}
\hline
\hline
\noalign{\smallskip}
Species X 	&  X/H$_2$ & Species X 	&  X/H$_2$\\	
\noalign{\smallskip}
\hline
\noalign{\smallskip}
C			& $5.82 \times 10^{-4}$	&	CO$_2$  		& $7.01 \times 10^{-5}$		\\
N   			& $1.60 \times 10^{-4}$    	&	CH$_3$OH  	& $1.40 \times 10^{-5}$		\\ 
O			& $1.16 \times 10^{-3}$	&	CH$_4$  		& $7.01 \times 10^{-6}$		\\
S        		& $3.66 \times 10^{-5}$	&	NH$_3$   		& $5.33 \times 10^{-5}$		\\
P			& $6.88 \times 10^{-7}$	&	N$_2$		& $5.33 \times 10^{-5}$		\\
H$_2$O  		& $5.15 \times 10^{-4}$	&	H$_2$S    	& $1.83 \times 10^{-5}$		\\
CO      		& $4.91 \times 10^{-4}$	&	PH$_3$		& $6.88 \times 10^{-7}$		\\ 
\hline
\end{tabular}
\end{center}
Elemental abundances derive from Lodders (2003). Molecular abundances are determined from our nominal gas phase composition.
\label{lodders}
\end{table}

\begin{figure}
\begin{center}
\resizebox{\hsize}{!}{\includegraphics[angle=0]{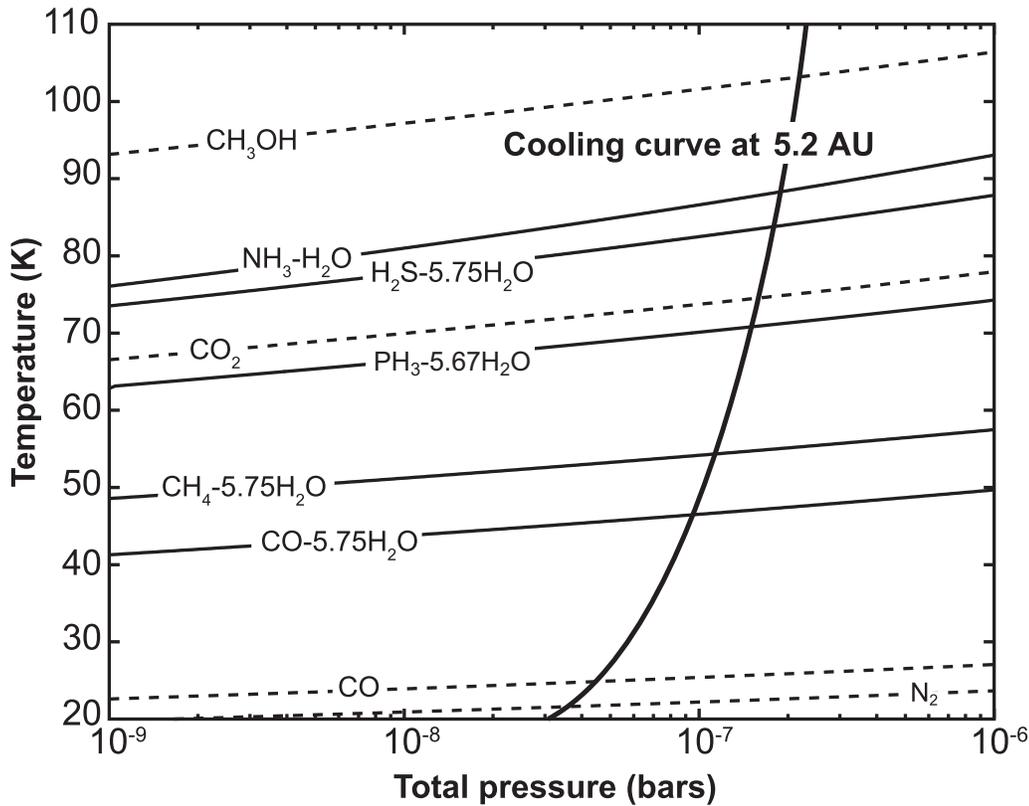}} \caption{Equilibrium curves of hydrate (NH$_3$-H$_2$O), clathrates (X-5.75H$_2$O or X-5.67H$_2$O) (solid lines), and pure condensates (dotted lines), and cooling curve of the protoplanetary disk at the distance to the star of 5.2 AU, assuming full (100\%) efficiency of clathration. Species remain in the gas phase above the equilibrium curves. Below, they are trapped as clathrates or simply condense.}
\end{center}
\label{cool1}
\end{figure}

In order to define the initial gas phase composition of the disk, we follow the approach of Mousis et al. (2009) and consider a gas phase in which all elements are in solar composition (Lodders 2003). Oxygen, carbon, and nitrogen are postulated to exist only in the form of H$_2$O, CO, CO$_2$, CH$_3$OH, CH$_4$, N$_2$, and NH$_3$. The abundances of CO, CO$_2$,  CH$_3$OH, CH$_4$, N$_2$ and NH$_3$ are then determined from the adopted CO/CO$_2$/CH$_3$OH/CH$_4$ and N$_2$/NH$_3$ gas phase molecular ratios. Once the abundances of these molecules are fixed, the remaining O gives the abundance of H$_2$O. We then set CO/CO$_2$/CH$_3$OH/CH$_4$~=~70/10/2/1 in the gas phase of the disk, values that are consistent with the ISM measurements considering the contributions of both gas and solid phases in the lines of sight (Marboeuf et al. 2008). In addition, S is assumed to exist in the form of H$_2$S, with H$_2$S/H$_2$ = 0.5 $\times$ (S/H$_2$)$_{\odot}$, and other refractory sulfide components (Pasek et al. 2005). We also consider N$_2$/NH$_3$ = 1/1 in the gas phase of the disk. This value is compatible with thermochemical calculations in the solar nebula that take into account catalytic effects of Fe grains on the kinetics of N$_2$ to NH$_3$ conversion (Fegley 2000). In the following, we adopt these mixing ratios as the nominal gas phase composition of our protoplanetary disk (see Table \ref{lodders}). The process by which volatiles are trapped in icy planetesimals, illustrated in Fig. \ref{cool1}, is calculated using the equilibrium curves of ammonia hydrates, clathrates and pure condensates, and the thermodynamic path detailing the evolution of temperature and pressure at the assumed distance to the star of 5.2 AU in the protoplanetary disk. This distance to the star has been arbitrarily chosen equal to that of Jupiter from the Sun as illustrative of the condensation sequence of the ices in the protoplanetary disk. The adoption of any other thermodynamic path located beyond the snow line would give the same information, provided that the temperature of the disk reaches values low enough to allow the formation of pure condensates at 20--30 K.

The equilibrium curves of hydrates and clathrates derive from Lunine \& Stevenson (1985)'s compilation of published experimental work, in which data are available at relatively low temperatures and pressures. On the other hand, the equilibrium curves of pure condensates used in our calculations derive from the compilation of laboratory data given in the CRC Handbook of Chemistry and Physics (Lide 2002). The cooling curve intercepts the equilibrium curves of the different ices at particular temperatures and pressures. For each ice considered, the domain of stability is the region located below its corresponding equilibrium curve. The clathration process stops when no more crystalline water ice is available to trap the volatile species. Note that, in the pressure conditions of the disk, CO$_2$ is the only species that crystallizes at a higher temperature than its associated clathrate. We then assume that solid CO$_2$ is the only existing condensed form of CO$_2$ in this environment. In addition, we have considered only the formation of pure ice of CH$_3$OH in our calculations since, to our best knowledge, no experimental data concerning the equilibrium curve of its associated clathrate have been reported in the literature. Figure \ref{cool1} illustrates the case where the clathration efficiency is total in the protoplanetary disk, implying that guest molecules had the time to diffuse through porous water-ice planetesimals before their accretion by proto-HD 189733b on its migration pathway. In this case, NH$_3$, H$_2$S, PH$_3$, CH$_4$ and $\sim$ 36\% of CO form NH$_3$-H$_2$O hydrate, H$_2$S-5.75H$_2$O, PH$_3$-5.67H$_2$O, CH$_4$-5.75H$_2$O and CO-5.75H$_2$O clathrates with the available water. The remaining CO and N$_2$, whose clathration normally occurs at lower temperatures, remain in the gas phase until the disk cools enough to allow their condensation in the form of pure ices. Once formed, clathrates and pure ices agglomerated and were presumably incorporated into icy planetesimals accreted by HD 189733b. In Sec. \ref{ab}, we calculate the abundances of volatiles in the envelope of HD 189733b as a function of the mass of ices incorporated in the planet on its migration pathway.

\section{Abundances of volatiles in HD 189733b}
\label{ab}
Following the approach described in Mousis et al. (2009), we have calculated the composition of ices incorporated in planetesimals formed in the protoplanetary disk for full clathration efficiency. This allowed us to reproduce the observed volatile abundances by adjusting the mass of ices vaporized in the envelope. To this end, two different approaches have been considered. We have first assumed that the O and C abundances measured by S09a are representative of the bulk  O and C composition of HD 189733b's envelope, since these values derive from measurements of the dominant O and C compounds (H$_2$O, CO, CO$_2$ and CH$_4$) at the sampled pressure levels ($\sim$ a few bars). We attempted to match simultaneously these O and C abundances to our model that predict the relative abundances of volatiles in the envelope. The masses so derived constitute the minimum total mass of ices needed to reproduce the observations. In a second approach, we have considered the mass of heavy elements ($\sim$ $35 \pm 15$ \Mearth) predicted by Guillot (2008) in HD 189783b from internal structure models. With an ice-to-rock ratio of 1 as is invoked for the composition of planetesimals produced in the outer solar nebula (Mousis et al. 2009), the minimum mass of ices accreted and dissolved in the envelope of HD 189783b should be equal to 10 \Mearth~if the core has eroded with time\footnote{In the case of Jupiter, interior models also predict that the current mass of its core can be as low as zero (Saumon \& Guillot 2004).}. We have then determined the corresponding elemental abundances that would result from a homogeneous mixing of these 10 \Mearth~of volatiles in the giant planet's envelope. 

Table 3 gives the abundances of volatiles in HD 189733b determined in each of the two cases. It shows that the elemental abundances calculated assuming the presence of 10 \Mearth~of ices vaporized in the envelope are more than 8 times higher than those fitted to the observations of S09a. Figure 2 represents the volatiles enrichments relative to solar elemental abundances (see Table 3) calculated in HD 189733b as a function of the mass of ices accreted and dissolved in the envelope. This figure shows that the predicted elemental mixing ratios are $\sim$ 0.15--0.3 times solar in the envelope of HD 189733b if they are fitted to the determinations of S09a. These subsolar elemental abundances in HD 189733b require the accretion of only $\sim$ 1.2 \Mearth~of ices (including $\sim$ 0.6 \Mearth~of water) in the envelope. On the other hand, our fit to the minimum mass of heavy elements predicted by Guillot (2008) gives elemental abundances that are $\sim$ 1.2--2.4 times oversolar in the envelope of HD189733b. Such oversolar elemental abundances require the presence of $\sim$ 5 \Mearth~of water over the 10 \Mearth~of ices vaporized in the envelope. The calculated oversolar enrichments are lower than those observed in Jupiter\footnote{C, N, S, P, Ar, Xe and Kr abundances are observed 1.6--5.8 times oversolar in the Jovian atmosphere (see Mousis et al. 2009 and references therein).} because the budget of ices needed in Jupiter's envelope to fit the observed volatiles enrichments under the same conditions ($\sim$ 18.6 \Mearth; see Mousis et al. 2009) is higher than the one used here. However, because the plausible mass range of heavy elements is quite large in HD 189733b, it  is also compatible with enrichment values higher than those observed in Jupiter (see Fig. 2).

\begin{figure}
\begin{center}
\resizebox{\hsize}{!}{\includegraphics[angle=0]{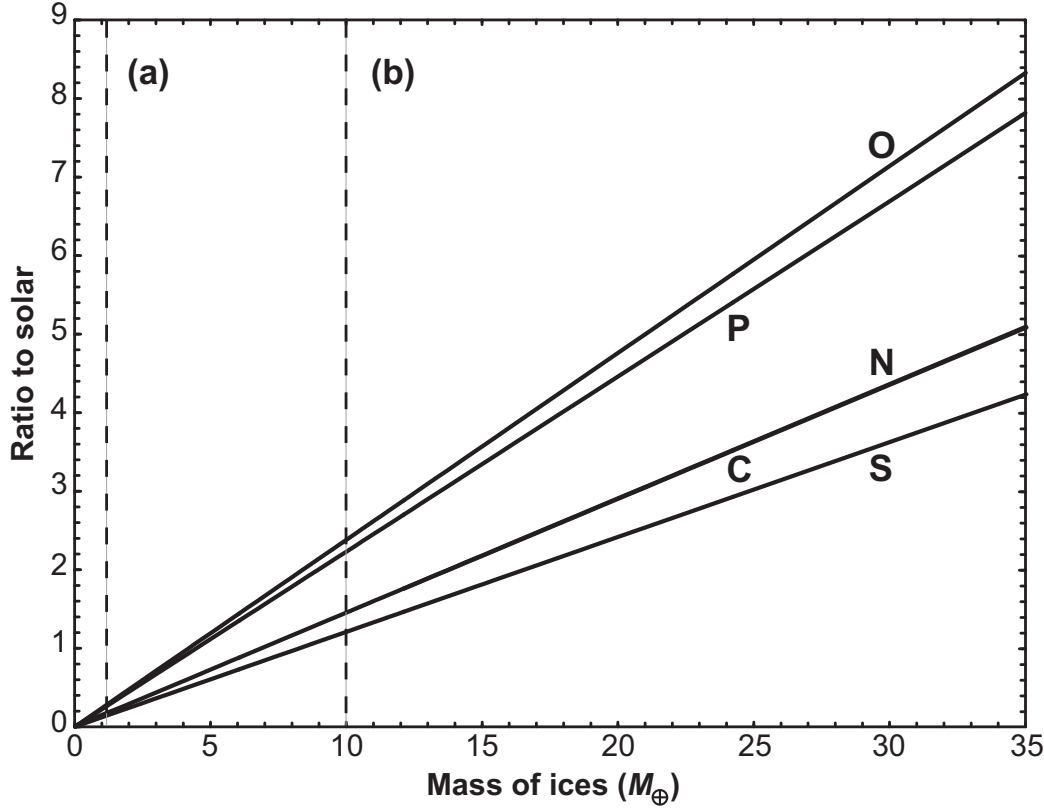}} \caption{Volatile enrichments in the envelope of HD 189733b as a function of the mass of accreted ices. N and C lines appear superimposed. The vertical lines corresponds to the enrichments in volatiles calculated in the case (a) of the simultaneous fit of the O and C abundances to the ones determined by S09a and (b) of the fit of the volatiles abundances to the minimum mass of heavy elements predicted by Guillot (2008).}
\end{center}
\label{enri2}
\end{figure}

\begin{table}
\caption[]{Calculated elemental abundances of volatiles in HD 189733b.}
\begin{center}
\begin{tabular}{lcc}
\hline
\hline
\noalign{\smallskip}
Element		 			& \multicolumn{2}{c}{Calculated abundance}			\\
Case					& (a)						& (b)					\\
\noalign{\smallskip}
\hline
\noalign{\smallskip}	
O						& $3.3 \times 10^{-4}$		&$2.8 \times 10^{-3}$ 	\\
C   				  	  	& $1.0 \times 10^{-4}$		&$8.4 \times 10^{-4}$  	\\
N   				    	 	& $2.8 \times 10^{-5}$		&$2.3 \times 10^{-4}$	\\
S   				    	    	& $5.3 \times 10^{-6}$		&$4.4 \times 10^{-5}$	\\
P						& $1.8 \times 10^{-7}$		&$1.5 \times 10^{-6}$	\\
\hline
\end{tabular}
\end{center}
Two different cases are considered: (a) simultaneous fit of the O and C abundances to the ones determined by S09a and (b) fit of the volatiles abundances to the minimum mass of heavy elements predicted by Guillot (2008).
\label{enri}
\end{table}

\section{Discussion}

The observation that HD189733b has subsolar abundances in its atmosphere poses some problems. Only $\sim$ 1 \Mearth~of ices must be accreted in the envelope to explain the observed C and O enrichments while internal structure models require the presence of $\sim$ 20-50 \Mearth~of heavy elements in HD189733b (Guillot 2008). This mass range of heavy elements should then be consistent with the observation of supersolar elemental abundances in the atmosphere, in a way similar to Jupiter, and not with the subsolar abundances observed by S09a. A number of explanations of this striking discrepancy may be advanced:

\begin {enumerate} 

\item {Planetesimals accreted by the growing planet did not ablate in the atmosphere but carried their volatiles intact  to the deep interior, where they remain. Simulations by Baraffe et al. (2006), based on the extended core-accretion models of Alibert et al. (2005a), show that 100 km planetesimals are destroyed in the envelope once the core mass reaches around 6 Earth masses, so this seems implausible.\\}

\item{ HD189733b started its formation at or beyond the ice-line and then migrated further to the inner parts of the disk, where planetesimals accreted by them were primarily rocky with little or no ice phase. While rocky planetesimals do contain varying amounts of volatiles, the most volatile species should be subsolar relative to the refractory phases. Were HD189733b to have migrated rapidly inward of the ice line, it is possible that most of the planetesimals it accreted were volatile poor. { However, formation models suggest that close-in giant planets   formed beyond the snow line before they started their migration (Lin et al. 1996; Fogg \& Nelson 2005, 2007; Mandell et al. 2007). Since dynamical models
show that the ratio of accreted to ejected planetesimals converges towards zero in the case of giant planets with masses reaching or exceeding that of Jupiter (Guillot \& Gladman 2000), the amount of planetesimals accreted by HD 189733b on its migration trajectory should have been very low. Enough material might have been accreted to be  compatible with the elemental subsolar abundances inferred by Swain et al. (2009), but would have been insufficient to explain the presence of 20-50 \Mearth~of heavy elements predicted by internal structure models (Guillot 2008).\\}

}

\item{The C and O atmospheric abundances may not be representative of the envelope composition, either because they are sequestered in species too refractory to be found in the atmosphere, or because the atmosphere is isolated from the interior by a radiative zone. The former seems implausible given the stability and observability of H$_2$O and CO as carbon and oxygen carriers. HD189733b is strongly irradiated, which leads to the development of an outer radiative zone extending down to the kilobar pressure level. We therefore hypothesize that differential settling, resulting from the combination of gravity and irradiation effects, took place inside the atmosphere of HD189733b, thus lowering the C and O abundances in the upper layers (Baraffe et al. 2009).}

\end{enumerate}

We find explanation (3) most plausible. In particular, the large heavy element abundance inferred in the deep interior by Guillot (2008) suggests that large amounts of material were condensed and available both prior to and during the collapse of the gaseous envelope to form the giant planet. For that reason as well, a scenario in which direct collapse formed the planet which was later seeded by heavy elements (Helled et al. 2006) would seem not to be applicable in the case of HD189733b. Furthermore, if our scenario of differential settling is correct, it should apply to all strongly irradiated extrasolar planets. Moreover, the N, S and P abundances calculated from the simultaneous fit of the O and C abundances to those observed by S09a may be tested by future observations of HD189733b's atmosphere. Elemental sulfur is expected to be contained mainly in H$_2$S at altitudes of 0.002--1 bar in HD189733b (Zahnle et al. 2009). Its abundance at these levels may be sampled provided observations at sufficiently high spectral resolution of its strong 2 $\mu$m~feature become available. On the other hand, the elemental abundances of N and P are likely more difficult to measure in HD189733b. The main nitrogen-bearing and phosphorus-bearing gases are predicted to be N$_2$ and P$_2$ (Visscher et al. 2006); both lack a dipole moment, and are therefore challenging to measure at IR wavelengths. In exoplanets that are cooler than HD189733b, elemental N and P are mainly in NH$_3$ and PH$_3$  below altitudes where photolysis occurs (Saumon et al. 2000; Visscher et al. 2006), and these molecular species are spectroscopically much easier to detect. 

\begin{acknowledgements}
O.M., G.T. and J.-P.B. acknowledge the financial support of the ANR HOLMES. A.P.S. acknowledges support from the NASA Origins program. G.T. is supported by Royal Society. J.I.L. is supported by JPL's Distinguished Visiting Scientist Program. Many thanks to an anonymous Referee whose comments have helped us to improve our manuscript.
\end{acknowledgements}

\end{document}